\documentclass[aip,graphicx]{revtex4-1}
\usepackage{graphicx}
\draft % marks overfull lines with a black rule on the right

\begin{document}

%\preprint{}

\title{Vernier effect within a versatile femtosecond optical parametric oscillator for broad-tunable, high-repetition-rate oscillation} %Title of paper



\author{Yuwei Jin}
\author{Simona M. Cristescu}
\author{Frans J. M. Harren}
\author{Julien Mandon}
%\thanks{}
\affiliation{Molecular and Laser Physics, Institute for Molecules and Materials, Radboud University, P.O. Box 9010, 6500 GL Nijmegen, The Netherlands}


\date{\today}

\begin{abstract}
\textbf{Within a synchronously pumped optical parametric oscillator (SPOPO), the inherent synchronism between the pump and the resonating signal is the magic to partly transfer the coherence property of the pump to the signal. In our demonstration, Vernier effect is observed within a femtosecond SPOPO by simply detuning the FSR of the cavity, generating signal pulses at tunable repetition rate from several GHz to 1~THz with a maximum 22.58~nm full width half maximum (FWHM) bandwidth supporting 160~fs pulses covering the C- and L-bands of the telecom wavelength region. This technique offers a simple method of active filtering of dense frequency comb lines instead of using Fabry-P\'{e}rot (FP) cavities with complex locking system for astro-comb generation. Beside, as a promising source for frequency combs with tunable and large comb-spacing, it offers potential opportunities for applications such as high speed coherent data transmission, line-by-line pulse shaping, optical clocks and precision metrology.}  
\end{abstract}


\maketitle %\maketitle must follow title, authors, abstract and \pacs

\noindent The generation of ultra-high repetition-rate coherent pulse trains plays an important role in a variety of applications including precision spectroscopy, ultrafast telecommunications, optical clocks and astronomical spectrograph calibration\cite{Janke2005,Quinlan2010,Li2008,Ycas2012,Wilken2012,Glenday2015,Steinmetz2008}. SESAM-mode-locked Er,Yb:galss lasers operate in the 1.5~$\mu$m spectral region have enabled increasing the repetition rate up to 100~GHz\cite{Oehler2010}. For a typical Ti:sapphire mode-locked laser, a repetition rate of 10~GHz was achieved\cite{Bartels2008}, and even higher repetition rate can be reached via using a Fabry-P\'{e}rot (FP) cavity after the laser source\cite{Chen2008,Kirchner2009}. The Fourier transform of such a high-repetition-rate pulse trains is defined as a frequency comb with large and equal comb-spacing as long as the laser has been actively stabilized\cite{Bartels2009,Heinecke2009}, and complex Pound-Drever-Hall (PDH) locking is also needed when a FP cavity is used as a comb-line filter for the repetition-rate multiplication\cite{Steinmetz2009,Li2012}. 

Nowadays, femtosecond optical parametric oscillators (fs-OPOs) are well established mid-infrared laser sources taking advantages of high output power, broad tuning range, and broad bandwidth\cite{Burr1997}. Due to the lack of frequency comb sources in the mid-infrared wavelength region beyond 2~$\mu$m, fs-OPO has been considered as a perfect candidate to realize high power mid-infrared optical frequency comb generation\cite{Gebs2008,Adler2009}. For a synchronously pumped fs-OPO, one can assume that the effective cavity length for fundamental pumping is $L_{f}$, maintaining a same repetition rate between the pump and the resonating signal. To generate high repetition rate signal pulse trains, harmonic pumping can be used, for which the effective OPO cavity length $L_{OPO}$ = $N/M\times L_{f}$ is chosen, where $N$ and $M$ are both integers, and $N/M$ is irreducible fraction ($N<M$). The fundamental repetition rate can thus be multiplied by a factor of $M$. Multi-GHz repetition rate oscillation based on harmonic pumping can be achieved for both singly-resonate and degenerate OPOs\cite{Jiang2002,Zhang2002,Lecomte2005,Kokabee2009,Vainio2012,Ingold2014}. In addition, high repetition rate optical frequency comb can also be achieved by actively stabilizing the OPO and coupling the signal into a FP cavity, resulting in a 10~GHz mode spacing in the optical frequency domain\cite{Zhang2015}.

\begin{figure}[!b]
	\centering
	\includegraphics[width=\linewidth]{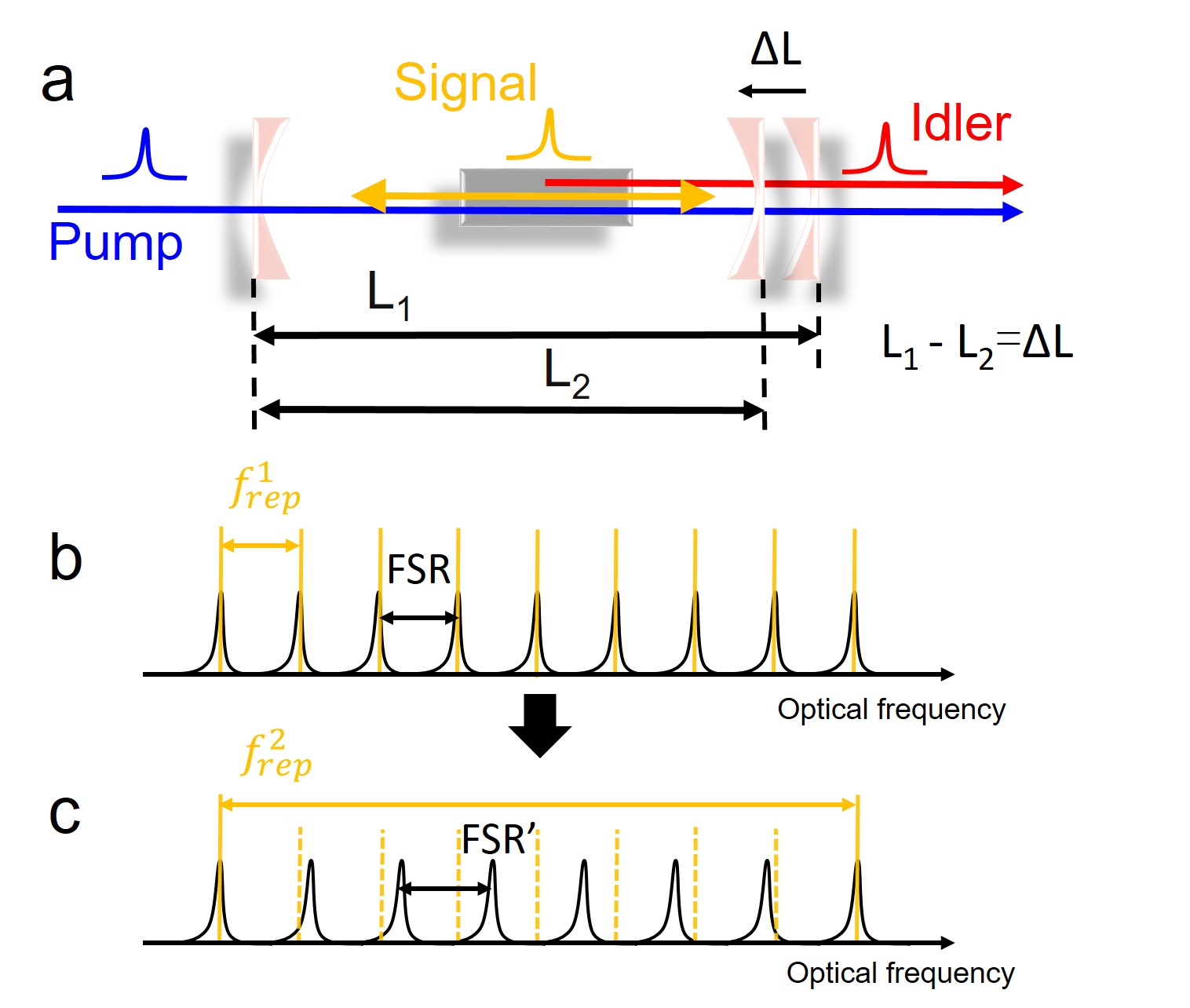}
	\caption{Basic principle of Vernier effect within a fs-OPO. \textbf{(a)}~Schematic of a fs-OPO, Vernier effect can be observed by detuning the cavity length. \textbf{(b)}~For fundamental synchronously pumping the signal comb lines are matching with the cavity modes. \textbf{(c)}~By detuning the cavity length, only part of the combs lines are matching with part of the cavity modes periodically along with the optical frequency axis. $\mathbf{f_{rep}^1}$, Fundamental repetition rate of the pump laser. \textbf{FSR}, Free spectral range of the OPO cavity. \textbf{FSR'}, Free spectral range for the length-detuned cavity. $\mathbf{f_{rep}^2}$, Comb spacing when the cavity length is detuned.}
	\label{fig:Veriner-effect-principle}
\end{figure}
For a fs-OPO, fundamental pumping scheme which is the most effective method can realize a record-high repetition rate of 10~GHz\cite{Lecomte2002}, limited mainly by the pump source. Meanwhile, Shorter cavity length is needed to achieve high repetition rate oscillation, limited by the lower efficiency due to the large beam waist size in the nonlinear crystal of a shorter cavity. For the FP cavity scheme, complex PDH locking is needed, determining a better transmission of selected comb lines. To avoid all the drawbacks mentioned above, we consider to use the harmonic pumping scheme with large numbers of previously defined $M$ and $N$, simply achieved by slightly detuning the cavity length. The similar technique was demonstrated previously in the picosecond regime, realizing a multi-gigahertz repetition rate oscillation\cite{Kimmelma2013}. In this letter, we report on a femtosecond laser pumped optical parametric oscillator based on MgO-doped periodically poled lithium niobate (MgO:PPLN) crystals, realizing high repetition rates tunable from multi GHz to more than 1~THz. In an extreme case, the fundamentally pumped signal pulse oscillates with the harmonically pumped signal pulse simultaneously. To our knowledge, this is the first demonstration of this phenomena for high repetition pulse generation in the femtosecond regime, and record-high repetition rate up to 1~THz are realized by simply detuning the cavity length. As most previous works interpreted this technique by using the term of ``harmonic pumping'' in time domain, we also offer a new perspective of ``Vernier effect'' that is a perfect interpretation of this technique in the frequency domain\cite{Gohle2007,Rutkowski2014,Hardy2012}.

\begin{figure}[!b]
	\centering
	\includegraphics[width=\linewidth]{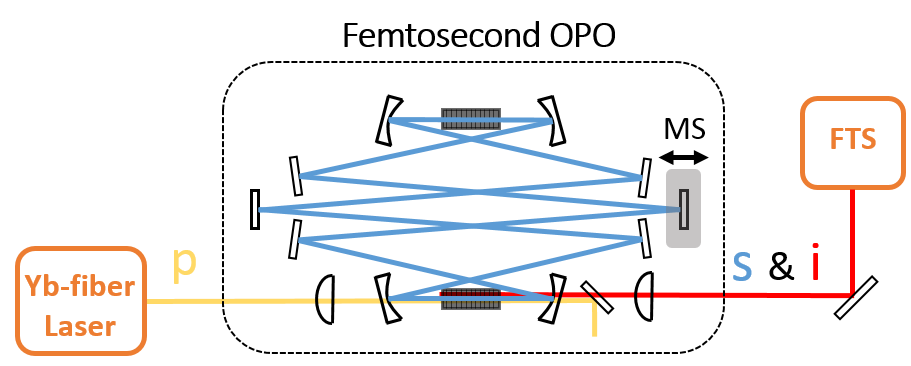}
	\caption{Experimental set-up of the fs-OPO pumped by a Yb-fiber laser, meanwhile a Fourier transform Spectrometer (FTS) is used to monitor the output of the OPO. \textbf{MS}: Moving stage. \textbf{p}, \textbf{s}, \textbf{i}: pump, signal and idler.}
	\label{fig:Set-up}
\end{figure}

The basic principle of Vernier effect within a fs-OPO is illustrated in Fig.~\ref{fig:Veriner-effect-principle}. For a fs-OPO, the pump is down-converted into the signal and the idler within a nonlinear crystal. By properly designing a cavity around the crystal, the signal can resonate inside the cavity, thus the idler is amplified by the different frequency generation (DFG) between the pump and the circulated signal. As can be seen in Fig.~\ref{fig:Veriner-effect-principle}(a), Vernier effect can be induced by slightly detuning the length of the OPO cavity, and the signal then travels more than one round trip to overlap with the pump again. For fundamental pumping, the resonating signal are excited within all cavity modes separated by a space equal to the repetition rate of the pump ($f_{rep}$), as illustrated in Fig.~\ref{fig:Veriner-effect-principle}(b). The free spectral range(FSR) of the cavity changes when the cavity length is detuned, the consequence is a periodically coupling between the fundamental signal comb lines and the modes of the cavity, as can be seen in Fig.~\ref{fig:Veriner-effect-principle}(c). In this case, only the comb lines within the modes of the cavity are excited, constituting a new comb structure with larger comb spacing corresponding to a higher repetition rate in time domain. To prove our theory, the experimental work described in this paper concentrates on the characterization of signal mainly in the frequency domain.     

\begin{figure}[!b]
	\centering
	\includegraphics[width=\linewidth]{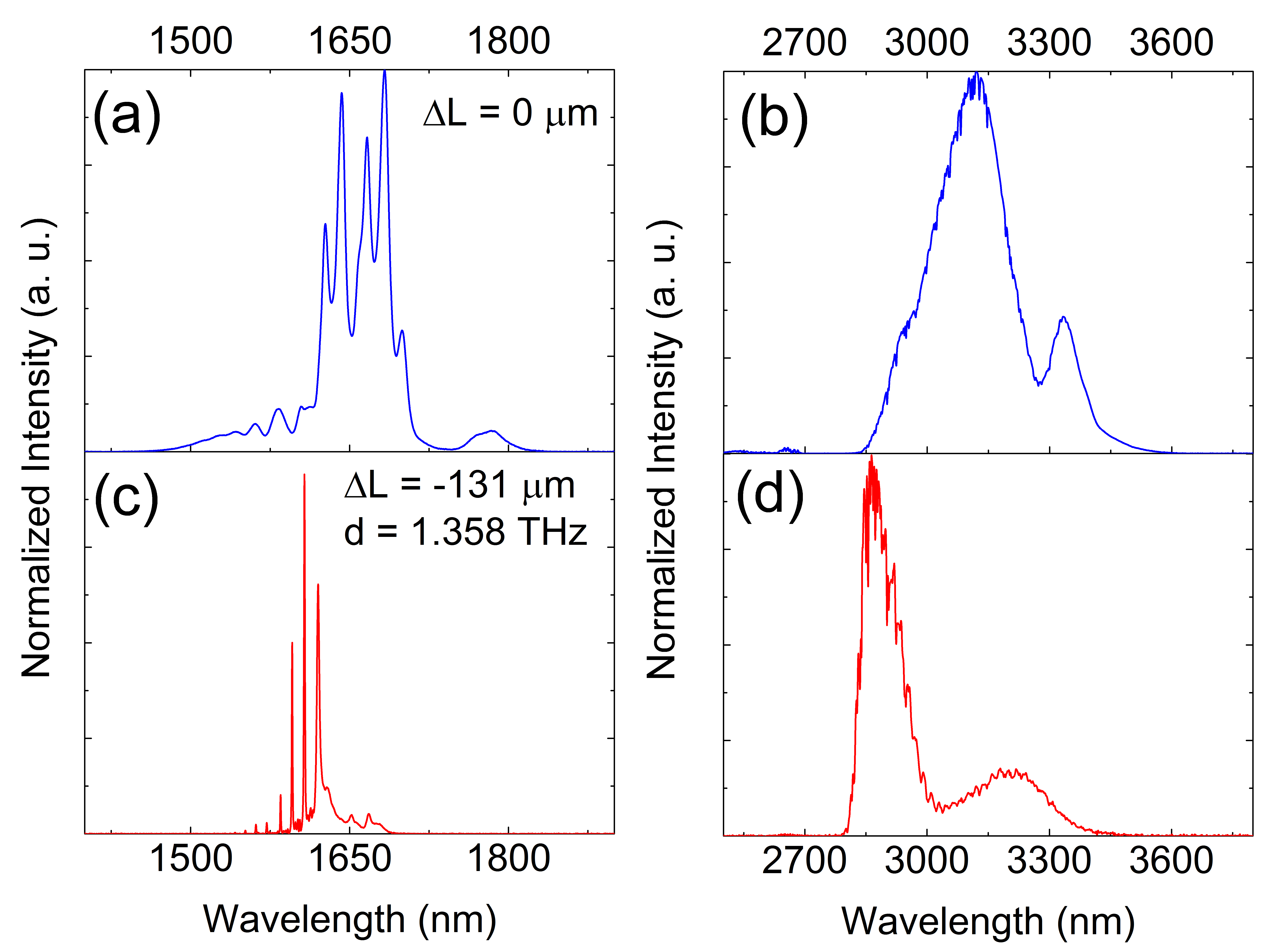}
	\caption{Normalized spectra of the OPO measured by the FTS at different cavity-length-detuning positions. \textbf{(a)}~and~\textbf{(b)}~Measured signal and idler spectra, respectively, when there is no detuning of the cavity length. \textbf{(c)}~and~\textbf{(d)}~Measured signal and idler spectra, when the detuning of the cavity length is -131~$\mu$m. \textbf{d}, distance between two comb structures.}
	\label{fig:Transition}
\end{figure}

\begin{figure}[!b]
	\centering
	\includegraphics[width=\linewidth]{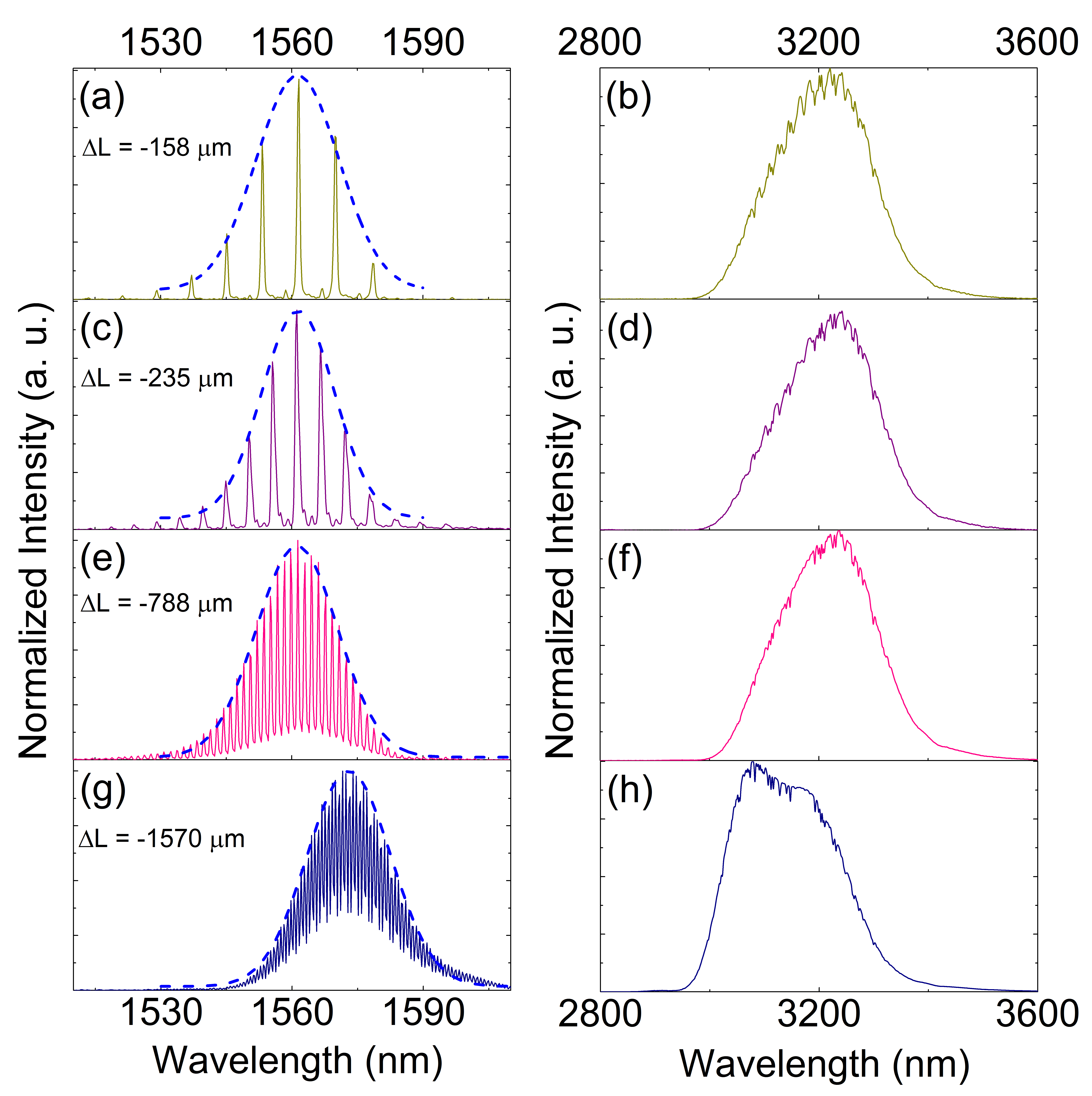}
	\caption{Normalized spectra of the signal(\textbf{a}, \textbf{c}, \textbf{e},  and \textbf{g}) and idler(\textbf{b}, \textbf{d}, \textbf{f},  and \textbf{h}) measured by the FTS for different cavity-length-detuning of -158, -235, -788 and -1570~~$\mu$m, respectively. The dotted blue curves are the Gaussian fittings.}
	\label{fig:Detuning_Spectra}
\end{figure}
The experimental set-up is shown in Fig.~\ref{fig:Set-up}. A fs-OPO is pumped by a high power Yb-fiber laser, and a commercial Fourier transform spectrometer (FTS) is used to monitor the output of the fs-OPO. The fs-OPO employed here is similar to that reported in \cite{Jin2014}. The difference is that only one crystal is pumped directly by a single Yb-fiber laser providing 80~fs pulses with an average power of 2~W. The total cavity length of the OPO is around 3.3~m in order to realize fundamental pumping at 90~MHz pump repetition rate. To achieve the Vernier effect, six chirped mirrors are used to compensate the dispersion mainly induced by the PPLN crystals with high reflectivity of 99.8~\% at the signal wavelength region. The zero dispersion wavelength is calculated to be at 1532~nm (6527~cm$^{-1}$), considering the chirped mirrors and the two 5~mm long crystals. 

The fundamental pumping synchronized by the pump repetition rate of 90~MHz is realized when the cavity length is around 3.3~m, the recorded signal and idler spectra are illustrated in Fig.~\ref{fig:Transition}(a) and Fig.~\ref{fig:Transition}(b), respectively. A spectral broadening of the signal can be observed, covering a full spectral range of 1000~nm centered at 1650~nm, which is mainly due to the cascaded quadratic nonlinearity\cite{Buryak2002}. As can be seen in Fig.~\ref{fig:Transition}(c) and Fig.~\ref{fig:Transition}(d),  when the cavity length is detuned by -131~$\mu$m, both fundamentally and harmonically pumped signals are resonating simultaneously within the OPO cavity, due to the cavity-length detuning tolerance of the fundamental pumping\cite{Jin2015}. In Fig. \ref{fig:Transition}(c), the spacing between two comb structures are 1.358~THz, corresponding to more than ten thousands round trips for one signal pulse to overlap with a pump pulse again. It is also worth noticing that, both fundamentally and harmonically pumped idlers (Fig.~\ref{fig:Transition}(b) and Fig.~\ref{fig:Transition}(d)) have the same repetition rate no matter if the cavity length is detuned, as the idler pulse can only be generated when the non-resonate pump presents\cite{Xu2015}.

\begin{table}[!t]
	\centering
	\caption{\bf Characterization of the cavity length detuning}
	\begin{tabular}{cccc}
		\hline
		$\Delta L$ ($\mu$m) & $f_{comb}$ (GHz) & FWHM (nm) & $\tau$ (fs) \\
		\hline
		-131  & 1358  & 27.69 & 138~fs \\
		-158  & 1040  & 22.42 & 160~fs \\
		-235  & 665.5 & 19.67 & 182~fs \\
		-581  & 260.2 & 20.21 & 177~fs \\
		-788  & 202.4 & 22.07 & 162~fs \\
		-1570 & 86.6  & 22.58 & 160~fs \\
		-3655 &       & 30.89 & 119~fs \\
		\hline
	\end{tabular}
	\label{tab:parameters}
\end{table} 

\begin{figure}[!b]
	\centering
	\includegraphics[width=\linewidth]{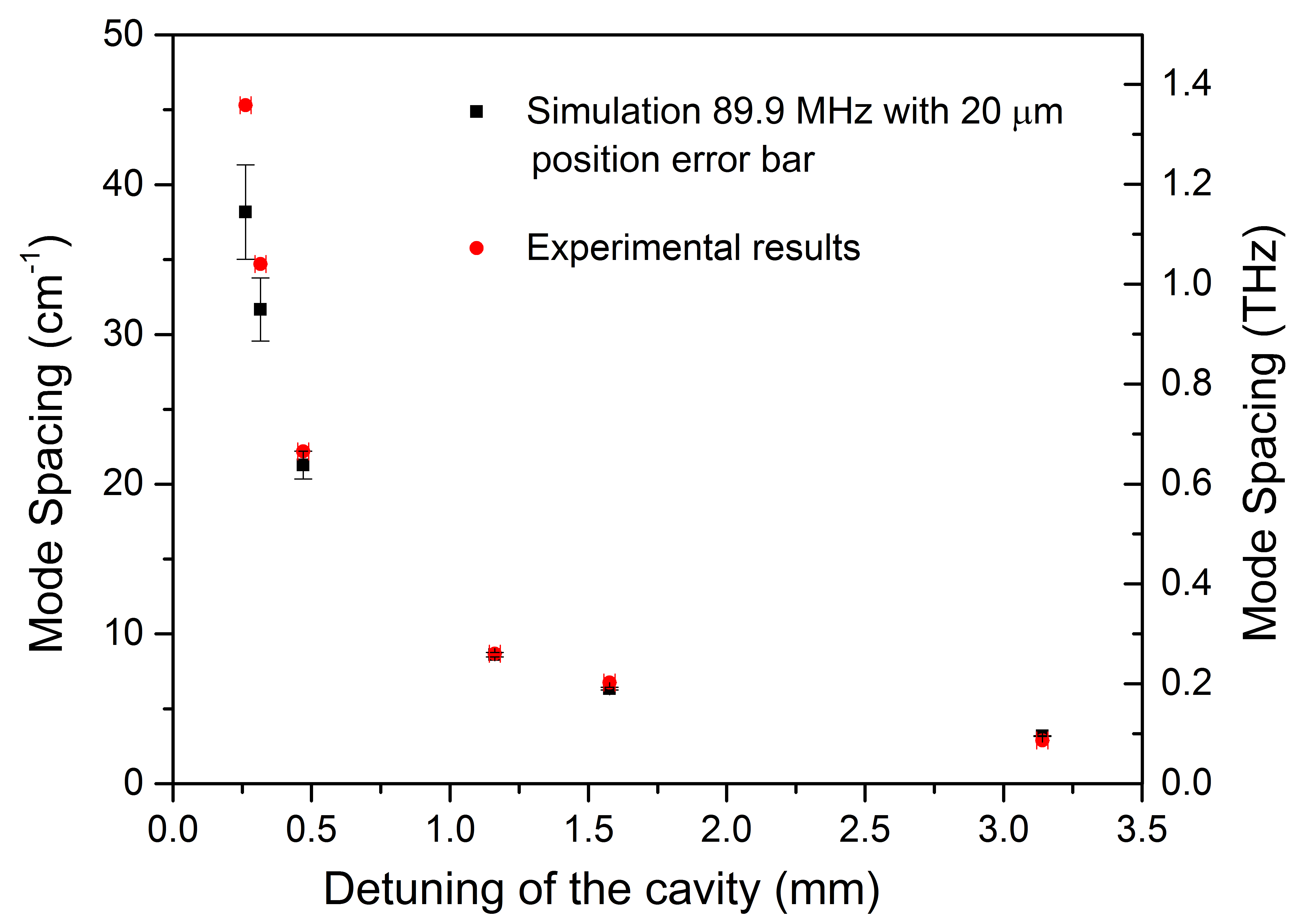}
	\caption{Comparison between the measured result and the simulation. For the simulation, an error bar considering a 20~$\mu$m position error is taken into account.}
	\label{fig:Meas-Sim}
\end{figure}
By further detuning the cavity length beyond the tolerance of the fundamental pumping, only harmonically pumped signal is able to resonate inside the OPO cavity at certain discrete detuning positions up to a detuned distance of -3655~$\mu$m. Some of the measured spectra for both signal and idler are shown in Fig.~\ref{fig:Detuning_Spectra}. Assume that the cavity length detuning is $\Delta L$, $f_{comb}$ represent the comb spacing of the signal, FWHM is the full width half maximum bandwidth and $\tau$ is the pulse duration in time domain. All parameters concerning the results shown in Fig. \ref{fig:Detuning_Spectra} are illustrated in Table \ref{tab:parameters}. For instance, when the cavity is detuned by -788~$\mu$m, the signal has a comb-like structure with a spacing of 202.5~GHz corresponding to the repetition rate of the signal. The FWHM bandwidth is 22.07~nm supporting 177~fs pulses. When the cavity-length-detuning is -3655~$\mu$m, the comb spacing can not be resolved by the FTS limited by its resolution.

A simulation has been done based on the perspective of the Vernier effect. Based on this assumption, the cavity length detuning can be calculated by
\begin{equation}
\Delta L = \frac{M-N}{M}\cdot L_f
\label{eq:detuninglength}
\end{equation} 
where $L_f$ is the cavity length for fundamental pumping, $M$ and $N$ are also previously defined. The comb spacing of the comb-like structure at different cavity-length-detuning position can then be describe as
\begin{equation}
f_{comb} = \frac{c\cdot(M-N)}{\Delta L}
\label{eq:fcomb}
\end{equation}
Where $c$ is speed of light. From Eq. \ref{eq:detuninglength} and Eq. \ref{eq:fcomb}, each array ($M$, $N$) can be mapped to an array of ($\Delta L$, $f_{comb}$), the relationship between $\Delta L$ and $f_{comb}$ can be solved. The simulation result is compared to the experimental result for the discrete different cavity-length-detuning positions as illustrated in Fig. \ref{fig:Meas-Sim}, indicating a good agreement between them. For the simulation, we take into account the real pumping repetition rate of 89.9~MHz and the tolerance of the cavity-length-detuning of the fundamental pumping cause the relative larger error when the detuning is near 0~$\mu$m.

In conclusion, we have demonstrated the Vernier effect in frequency domain within a synchronously pumped fs-OPO, that has been interpreted previously as harmonically pumping in time domain. The repetition rate of the signal can be tuned from multi-GHz to above 1~THz by slightly detuning the cavity length. This observation offers a simple method for the active filtering of dense frequency comb lines instead of using FP cavities with complex locking system to reach tens of GHz comb spacing. Compare to the previous work\cite{Kimmelma2013}, we extend this technique to fs-OPOs and give a interpretation in frequency domain linked to the well-known Vernier effect for the first time. The simulation gives good agreement compared to the experimental results. We believe that this tunable OPO will be an ideal source for astro-comb generation in the future by means of active stabilization as well as for ultrafast telecom applications.

\section*{Funding Information}

\textbf{Funding.} Technology Foundation STW (11830).

\section*{Acknowledgments}

\textbf{Acknowledgment.} The authors thank Dr. Lucile Rutkowski from Ume\aa{} university in Sweden for useful discussion.
 
\end{document}